\documentclass[superscriptaddress,%
 amsmath,amssymb,
preprint,%
]{revtex4-1}

\usepackage{graphicx}
\usepackage{dcolumn}
\usepackage{bm}
\usepackage{hyperref}

\begin{document}

\title{Reset and switch protocols at Landauer limit in a graphene buckled ribbon}

\author{I. Neri}
\affiliation{ 
NiPS Laboratory, Physics Department, Universit\`a degli studi di Perugia, via Pascoli, I-06123 Perugia, Italy.
}%
 \affiliation{INFN Sezione di Perugia, via Pascoli, I-06123 Perugia, Italy.}
\author{M. Lopez-Suarez}%
 \email{miquel.lopez@nipslab.org}
\affiliation{ 
NiPS Laboratory, Physics Department, Universit\`a degli studi di Perugia, via Pascoli, I-06123 Perugia, Italy.
}%
\author{D. Chiuchiù}%
\affiliation{ 
NiPS Laboratory, Physics Department, Universit\`a degli studi di Perugia, via Pascoli, I-06123 Perugia, Italy.
}%
\author{L. Gammaitoni}%
\affiliation{ 
NiPS Laboratory, Physics Department, Universit\`a degli studi di Perugia, via Pascoli, I-06123 Perugia, Italy.
}%

\begin{abstract}
Heat produced during a reset operation is meant to show a fundamental bound known as Landauer limit, while simple switch operations have an expected minimum amount of produced heat equals to zero. However, in both cases, present day technology realizations dissipate far beyond these theoretical limits. In this paper we present a study based on molecular dynamics simulations, where reset and switch protocols are applied on a graphene buckled ribbon, employed here as a nano electromechanical switch working at the thermodynamic limit.
\end{abstract}

\maketitle

\section{Introduction}
Mechanical switches have attracted the attention of the scientific community for a long time \cite{cha2005fabrication,fujita20073,jang2008nems,jang2008mechanically}. 
One interesting aspect is the non-volatile nature of bits encoded into bistable mechanical systems. This condition could in principle avoid the typical heat production associated with losses due to electric currents. In fact, excess heat production during computation is one of the main limitations that standard CMOS technology is presently facing in order to develop next generations of switches. Present CMOS switches are orders of magnitude far from the theoretical minimum heat required for computing. This fundamental bound, known as Landauer limit\cite{landauer}, arises for any physical system encoding two logic states, $0$ and $1$, and in contact with a heat bath at temperature $T$, when we intend to set the system into one given state, regardless of the knowledge of the initial state. According to the so-called Landauer principle, such a reset operation must produce at least $Q_L=k_BT\log2$ of heat (with $k_B$ Boltzmann constant)\cite{landauer}. However the Landauer limit does not close the door to switch procedures that do not generate heat, once the initial state is known and a proper switch protocol is observed\cite{ICT-Energy-Gamma}.

\begin{figure}[ht!]
\centering
\includegraphics[width=0.85\textwidth]{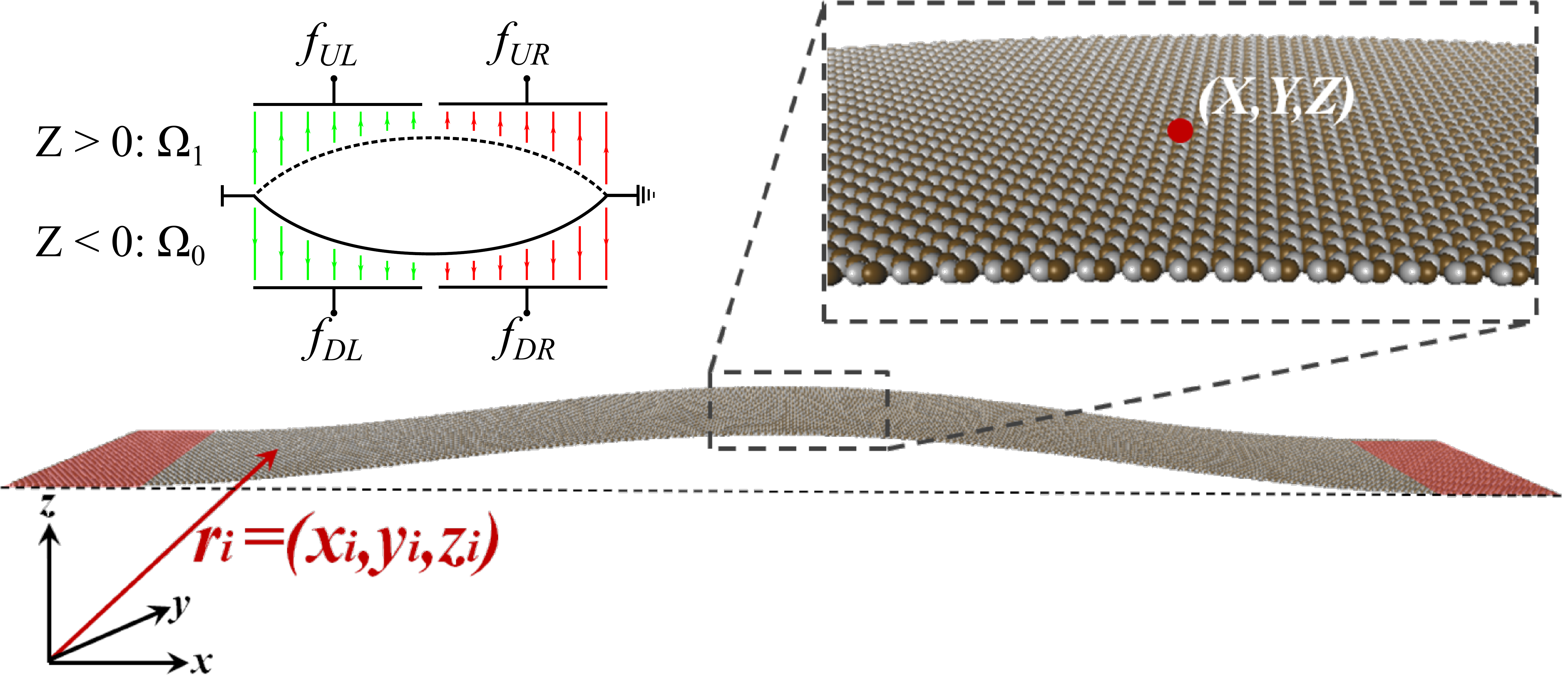}
\caption{Schematics of the 2D graphene ribbon switch encoding 1 bit of information. Top left diagram shows the two stable positions and the four electrodes used to actuate the selected protocol.}
\label{fig:scheme}
\end{figure}

In this paper we consider the encoding of a logical bit in a given configuration of a clamped-clamped single-layer graphene ribbon, as depicted in \autoref{fig:scheme}. To achieve two stable states, a compression is applied by bringing closer the extremes of the ribbon in the $x$-direction that buckles the structure in the out-of-plane direction. 
This compression generates an energy barrier separating the two stable states corresponding to the upward and downward buckling. To identify which logic state the ribbon represents at time $t$, we monitor the coordinates of the central atom in the ribbon (see zoomed image in \autoref{fig:scheme}). Being $(X,Y,Z)$ the coordinates of the central atom, we indicate with $\Omega_1$ ($\Omega_0$) the set of possible physical microstates with $Z>0$ ($Z<0$). If the system is a micro state that belongs to $\Omega_0$ ($\Omega_1$) we say that the ribbon encodes the logical state $0$ ($1$). Starting from these considerations we present a \emph{reset protocol} based on electrostatic forces acting on the graphene ribbon that approaches the Landauer limit. The procedure is designed to reach the theoretical thermodynamic bound without altering the compression of the ribbon. We also discuss a similar \emph{switch protocol} that realizes the switch process approaching $Q_{SW}=0$.

\section{Methodology}
Heat production during protocol execution is evaluated by using molecular dynamics simulations, performed with LAMMPS\cite{lammps}.
Here we consider a simulated ribbon consisting in a $6\times1~nm^2$ suspended structure composed by 240 carbon atoms.
Carbon-carbon interactions are described through the REBO potential\cite{brenner2002second}. The optimized lattice constant is $a=2.42~\AA$ which is in good agreement with both \emph{ab initio} calculations and experimental values\cite{graphene}.
The obtained Young modulus, $Y=0.85~TPa$, is also close to the expected value \cite{lee2008measurement}.
All the simulations are performed at temperature $T=10~K$. Electrostatic forces are applied on the structure by a set of four electrodes, classified by labels \emph{U (Up)}, \emph{D (Down)}, \emph{L (Left)} and \emph{R (Right)} as shown in the top left side of \autoref{fig:scheme}. As the graphene ribbon is considered to be electrically connected to the ground, $V_{graphene}=0$, the force produced by any of the electrodes is attractive if $V_{electrode}\neq0$ and zero otherwise. In the reference scheme, the \emph{Up} (\emph{Down}) electrodes generate positive (negative) forces along \textit{z} direction. 

The total energy of the graphene ribbon under the action of electric external fields can be written as 
\begin{equation}\label{eq:totalenergy}
H(\mathbf{P}, \mathbf{R}, t)= H_{kin}(\mathbf{P})+H_{int}(\mathbf{R})+H_{ext}(\mathbf{R},t)
\end{equation}
where $\mathbf{P}=\{\mathbf{p}_i\}$ and $\mathbf{R}=\{\mathbf{r}_i\}$ are the momenta and positions of all carbon atoms respectively. $H_{kin}(\mathbf{P})$ is the total kinetic energy of the ribbon, $H_{int}(\mathbf{R})$ the interatomic interaction energy and $H_{ext}(\mathbf{R},t)$ the total interaction energy between atoms and the external electric fields. In the case under consideration we have:
\begin{equation}\label{eq:protocol}
\begin{aligned}
H_{ext}(\mathbf{R},t)=\sum_{i=1}^{n} &\bigg[ \theta\left(x_i-\tfrac{l}{2}\right)
\left(\frac{f_{UL}(t)}{(g-z_i)}-\frac{f_{DL}(t)}{(g+z_i)}\right)+\\&+\theta\left(\tfrac{l}{2}-x_i\right)\left(\frac{f_{UR}(t)}{(g-z_i)}-\frac{f_{DR}(t)}{(g+z_i)}\right)\bigg]
\end{aligned}
\end{equation}
where $n$ is the number of atoms, $l$ is the clamp-clamp distance, $\mathbf{r}_i=\{x_i,y_i,z_i\}$, $\theta$ is the Heaviside function, and $g=10~nm$ is the distance between the electrodes and the plane $z=0$. Parameters $f_{UL}$, $f_{DL}$, $f_{UR}$ and $f_{DR}$ are relative to the corresponding electrode and their values depend on the voltages applied according to the given protocol.

Since the system is symmetric with respect to the $xy$ plane, both $0$ and $1$ logic states have the same average potential energy. This implies that quasi-static transformations between two logic states are characterized by a zero change in the average total energy: $\Delta H=0$, and consequently $Q=W$. In practice, transformations occurs in finite times, so $\Delta H$ may be nonzero. Since we consider protocol duration much larger than the relaxation time of the system ($\sim10~ps$), we make the assumption that $\Delta H=0$ also in practice. Under these conditions we can compute the generated heat as the average work performed on the system, via the Stratonovich integral\cite{seifert}:
\begin{equation}\label{eq:work}
W=\bigg\langle\int_{t_0}^{t_{end}} \frac{\partial H_{ext}(\mathbf{R},t)}{\partial t} d t \bigg\rangle
\end{equation} 
where $\langle \cdot \rangle$ denote averages over an ensemble of protocol realizations and $t_0$ and $t_{end}$ are the starting and ending time of the protocol.

\section{Normal mode analysis}
\begin{figure}[ht!]
\centering
\includegraphics[width=0.85\textwidth]{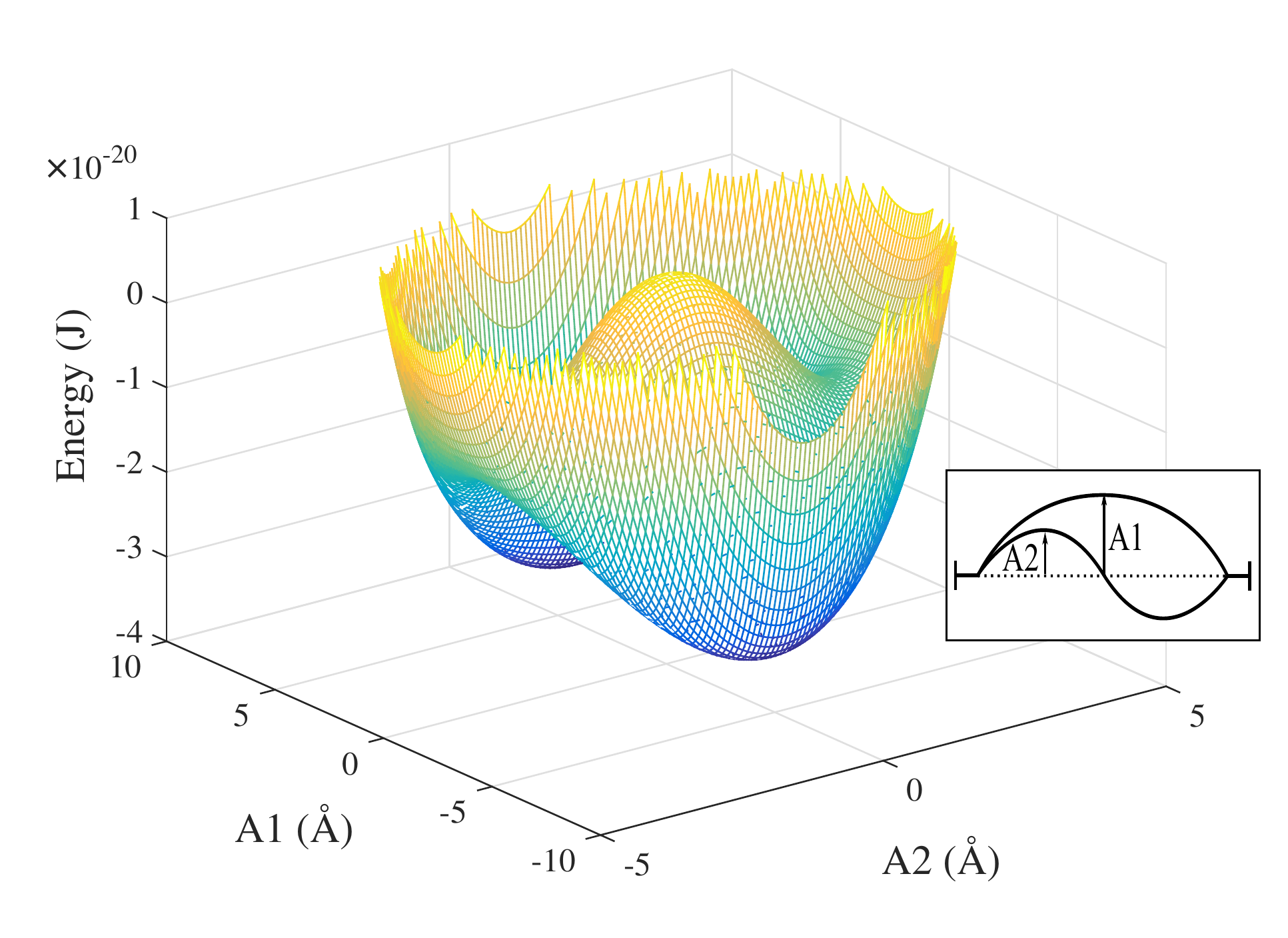}
\caption{Energy potential landscape $U(A_1,A_2)$ as function of the first two flexural mode amplitudes $A_1$ and $A_2$.  Inset: schematics of the two normal modes and their amplitudes.}
\label{fig:potential_landscape}
\end{figure}

In order to provide a visual description of the system dynamics, we focus our attention on the time evolution of the first two normal modes of the ribbon. Being $A_1$ and $A_2$ the amplitudes of the first two flexural modes, we can write an effective potential energy of the ribbon $U(A_1,A_2)$ in terms of these amplitudes (\autoref{fig:potential_landscape}). The local maximum in the potential landscape, $U(0,0)$, corresponds to a flat configuration of the ribbon. 

Trajectories on the landscape represent the evolution of the system undergoing a physical transformation. In particular we are interested in trajectories that go from one well to the other: they represent the physical realization of the switch between the two logic states. 
A trajectory that goes directly through the maximum at $U(0,0)$ corresponds to a \emph{snapping} of the ribbon. On the other hand trajectories that go around the maximum correspond to a dynamic that implies an intermediate \emph{S}-shape ribbon configuration. As a matter of fact, snapping trajectories can be hardly associated with a thermodynamically reversible (i.e. adiabatic) transformation due to the difficulty to control the velocity of the switch dynamics. On the other hand, \emph{S}-shape trajectories can be followed with an arbitrarily low velocity and thus result more promising for minimizing the heat production during the switch protocols.
Due to the presence of fluctuations the trajectories are random paths and are best monitored by computing the time evolution of the probability density function (PDF) of $A_1$ and $A_2$.

\section{Reset protocol}
\begin{figure}[ht!]
\centering
\includegraphics[width=0.85\textwidth]{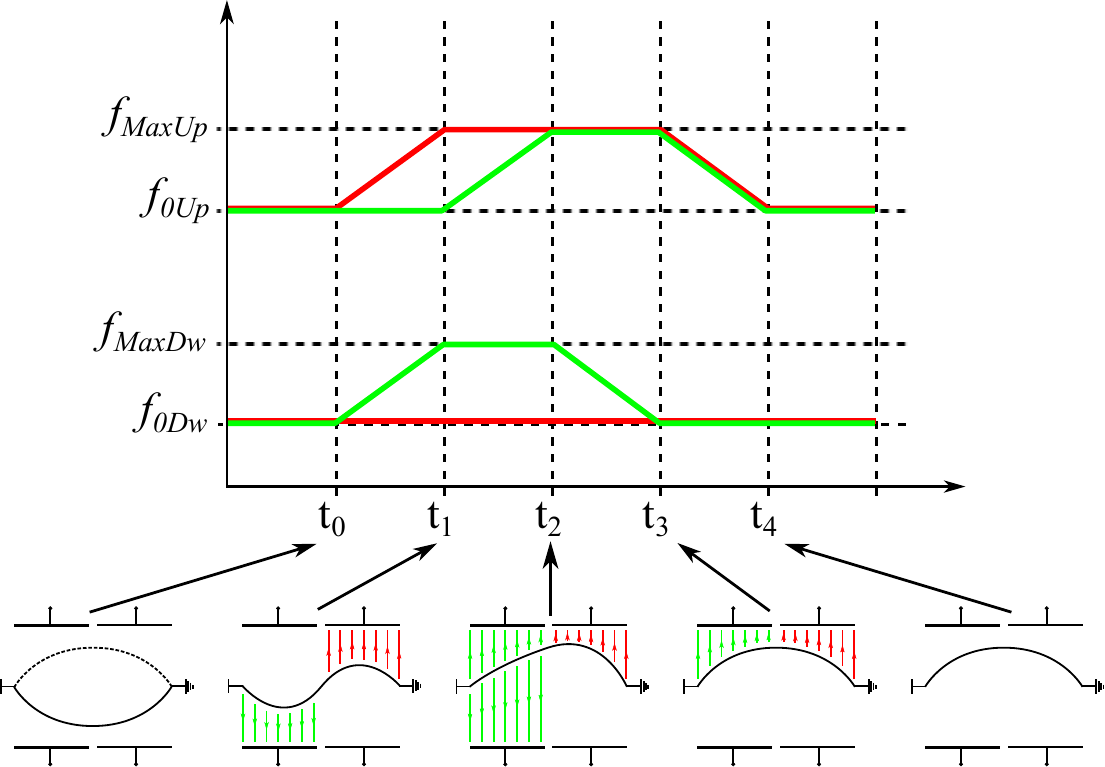}
\caption{Reset protocol.  Time diagram of the electrostatic forces acting on the ribbon. Top curves are relative to the top electrodes while the bottom ones are relative to the bottom electrodes; green and red colors represent left and right electrodes respectively. Lower panels represents the ribbon profile at different time.}
\label{fig:protocol_reset}
\end{figure}

In the following, we focus our attention on the reset operation. Such an operation is generally performed to bring to a known logic state a bistable system that is in an unknown state. This implies that at the beginning the system has $50\%$ of probability of being both in $\Omega_0$ and $\Omega_1$; after the reset operation the system ends with $100\%$ of probability in $\Omega_1$ (or equivalently in $\Omega_0$). In our case, the initial condition is prepared by setting $Z<0$ or $Z>0$ with $50\%$ of probability, and allowing the system to thermalize near the corresponding energy minimum, similarly to what is done in Ref. \cite{ciliberto}.
To perform the reset operation we apply the following protocol: a set of electrostatic forces (that can be expressed using \autoref{eq:protocol}) acting on each individual atom is applied according to the sequence presented in \autoref{fig:protocol_reset}.

\begin{figure}[ht!]
\centering
\includegraphics[width=0.85\textwidth]{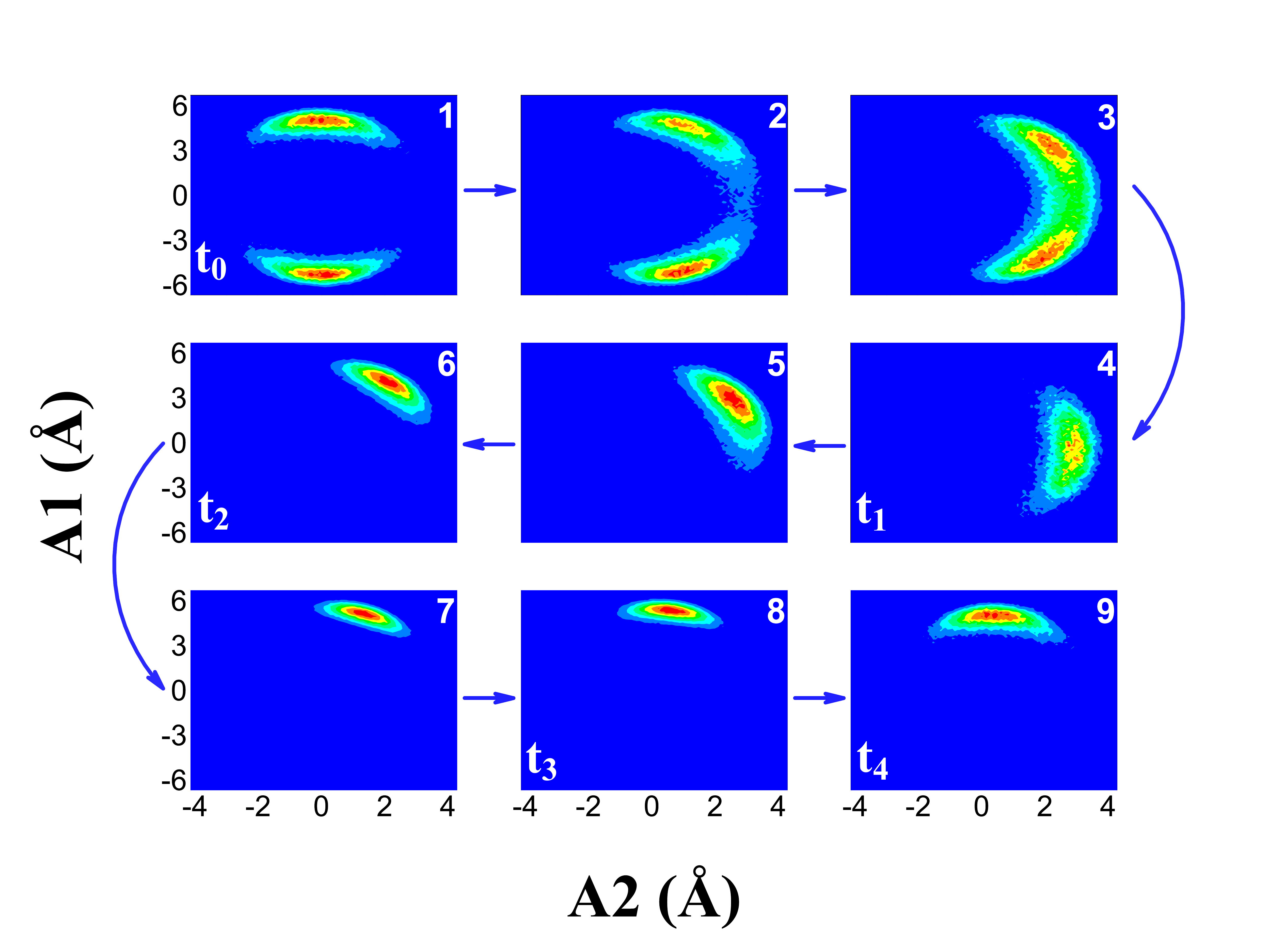}
\caption{Reset protocol. Evolution of the PDF of the first two flexural modes $A_1$ and $A_2$ for the reset procedure. Panels are numbered from 1 to 9 according with the time evolution indicated by the arrows.}
\label{fig:reset_potential_landscape}
\end{figure}

Applying the reset protocol, the PDF of $A_1$ and $A_2$ change as illustrated in \autoref{fig:reset_potential_landscape}. From time $t_0$ to $t_1$ (panels 1 to 4) the system evolves into a \textit{S}-shape ($\langle A_1\rangle=0$). From $t_1$ to $t_3$ (panels 4 to 7) the system goes from a \emph{S}-shape to a buckled ($\langle A_2\rangle =0$) configuration. At time $t_4$ the system is stable in the final logic state, the forces are removed and it thermalizes toward local equilibrium (\autoref{fig:reset_potential_landscape} panel 9).

\begin{figure}[ht!]
\centering
\includegraphics[width=0.85\textwidth]{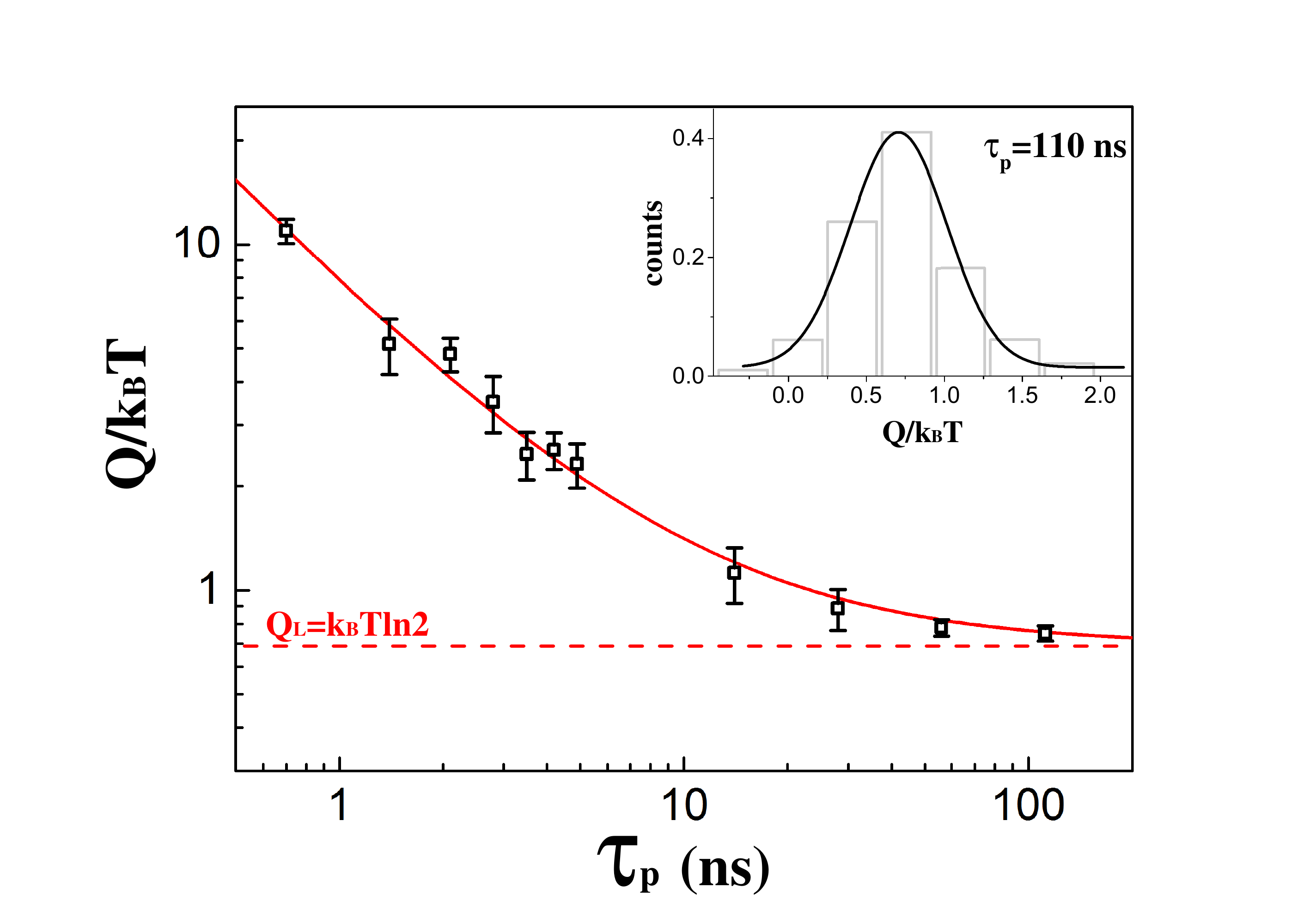}
\caption{Heat produced by the reset operation as function of the total protocol time $\tau_p$. The heat produced, obtained from the simulated experiment (squares), is plotted together with a fit to the data (solid line). The dashed line represents the Landauer limit. Inset: distribution of produced heat for $\tau_p=110~ns$.}
\label{fig:dissipation_reset}
\end{figure}

The total heat produced during the reset is computed by evaluating \autoref{eq:work} over the time interval $t_0-t_4$. In \autoref{fig:dissipation_reset} the produced heat is represented as function of the total time of the protocol $\tau_p$. Continuous line represents the fit with $Q=k\tau_p^{-1}+Q_L$. For large values of $\tau_p$ the total produced heat approaches $k_BT\log2$ as prescribed for a total entropy reduction of $k_B\log2$\cite{landauer}. 

It's interesting to note that the trend $Q=k\tau_p^{-1}+Q_L$ is empirical and, to be consistent with the literature, must be lower bounded by the limit $\kappa_{min}\tau_p^{-1}+Q_L$ derived in \cite{aurell}. This requires that $\kappa_{min }\leq k$ with equality saturated only if our protocol is proven to be optimal in the sense of \cite{aurell}. This study goes beyond the scope of the present paper and will not be discussed here.

\section{Switch protocol}
We now focus our attention on the switch operation. A switch operation consists in a transformation that brings the system from a known logic state to a known (and different) logic state (e.g. from $\Omega_0$ to $\Omega_1$ or \emph{vice-versa}). For the sake of discussion, in the following we will consider the transformation from the state $\Omega_0$ to $\Omega_1$. In this case, at difference with the reset protocol, being the total entropy variation equal to zero, the expected minimum heat produced by the procedure is $Q_{SW} = 0$.

\begin{figure}[ht!]
\centering
\includegraphics[width=0.85\textwidth]{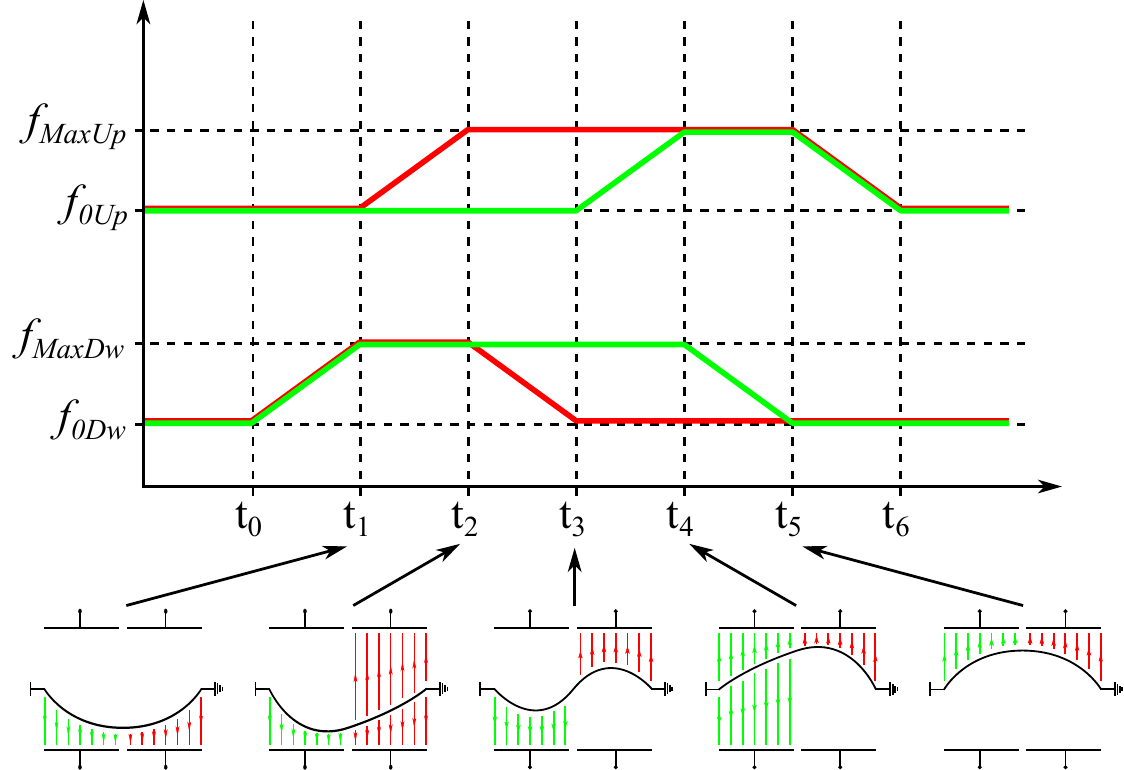}
\caption{Switch protocol.  Time diagram of the electrostatic forces acting on the ribbon as in \autoref{fig:protocol_reset}.}
\label{fig:protocol_switch}
\end{figure}

\begin{figure}[ht!]
\centering
\includegraphics[width=0.85\textwidth]{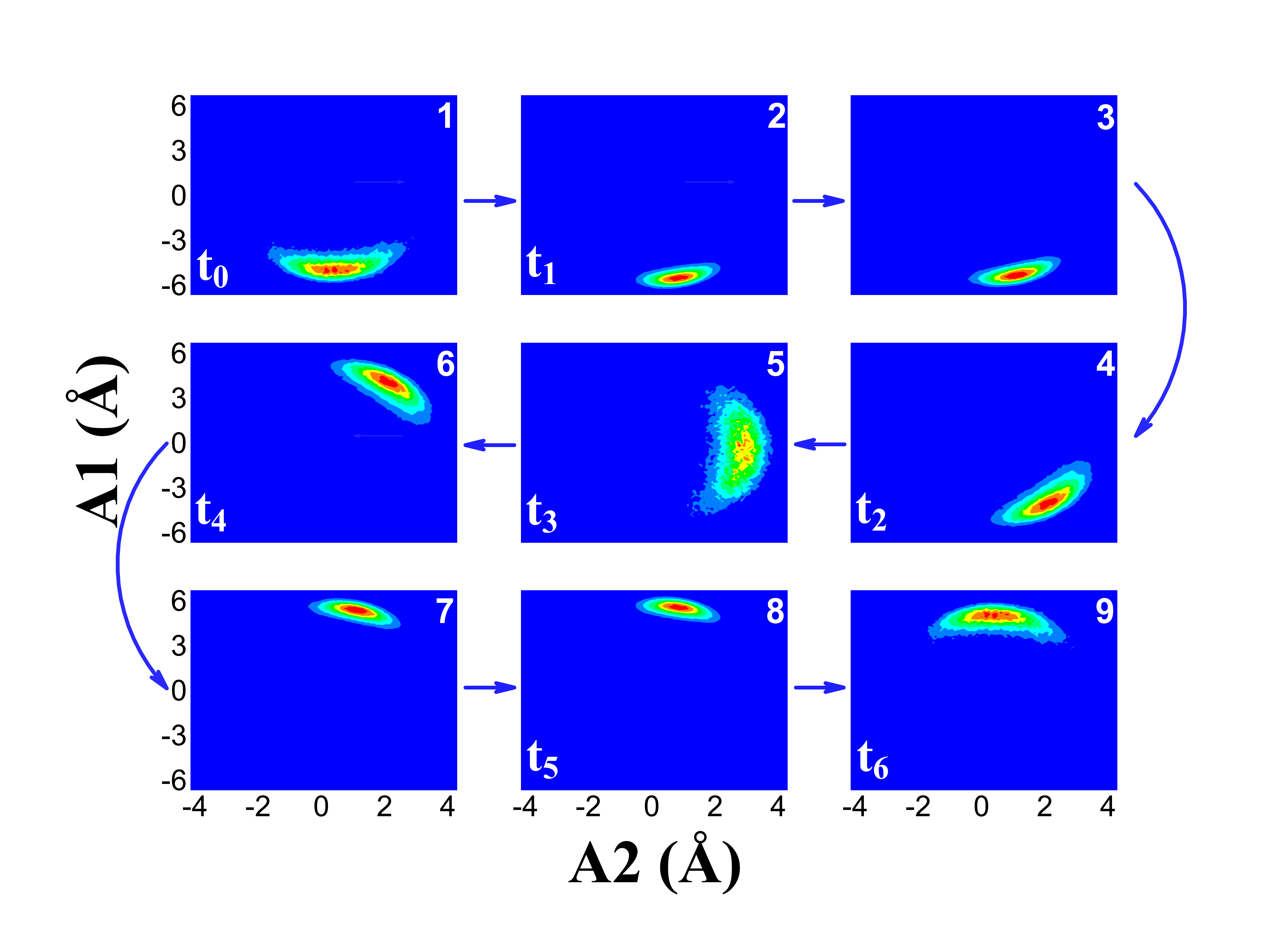}
\caption{Switch protocol. Evolution of the PDF of the first two flexural modes $A_1$ and $A_2$ for the reset procedure. Panels are numbered from 1 to 9 according with the time evolution indicated by the arrows.}
\label{fig:switch_potential_landscape}
\end{figure}

The switch protocol relative to \autoref{eq:protocol} is presented in \autoref{fig:protocol_switch}. The system is considered to be initially in local equilibrium in $\Omega_0$. In the first time frame ($t_0-t_1$) it is in the buckled configuration by applying a large electrostatic force (\autoref{fig:switch_potential_landscape} panel 2). In the following time frames ($t_1-t_3$) the ribbon is gently changed in a \emph{S}-shape configuration ($\langle A_1\rangle=0$, \autoref{fig:switch_potential_landscape} panel 3-5). Subsequently the system evolves into a shape with $\langle A_2\rangle=0$, confining the system in the destination state (\autoref{fig:switch_potential_landscape} panel 8). Finally, at time $t_6$ all the forces are removed and the system relaxes at local equilibrium in the final configuration (\autoref{fig:switch_potential_landscape} panel 9).

\begin{figure}[ht!]
\centering
\includegraphics[width=0.85\textwidth]{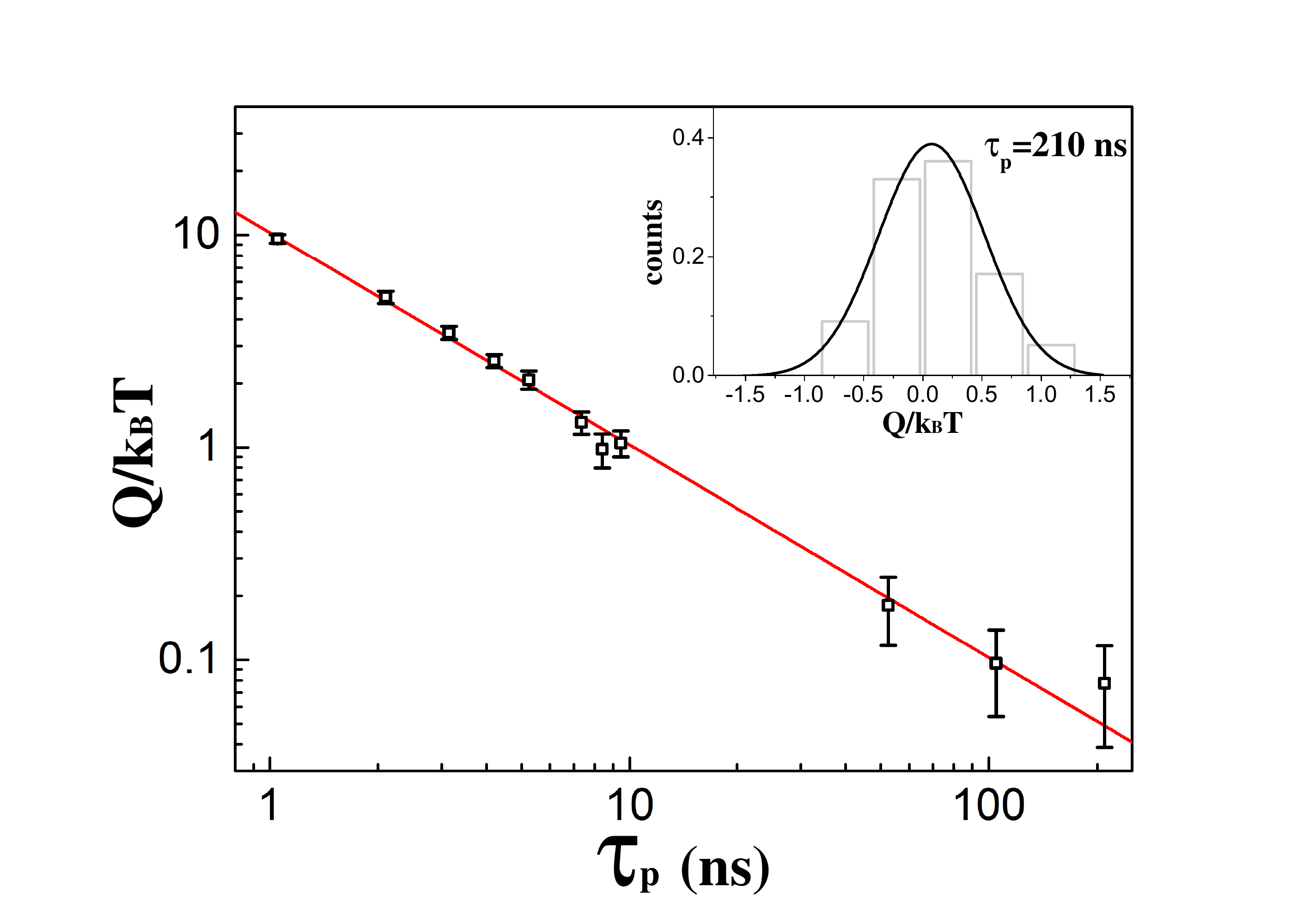}
\caption{Heat produced by the switch operation as function of the total protocol time $\tau_p$. Produced heat approaches zero by increasing $\tau_p$. The heat produced, obtained from the simulated experiment (squares), is plotted together with a fit to the data (solid line). Inset: distribution of produced heat for $\tau_p=210~ns$. }
\label{fig:dissipation_soft}
\end{figure}

As before, the work done on the system is evaluated according to \autoref{eq:work}. Results are presented in \autoref{fig:dissipation_soft} where the expected value $Q_{SW}=0$ is asymptotically approached as the duration of the protocol increases. The produced heat follows the trend $Q=k\tau_p^{-1}$  (solid line). For slow protocols, approaching the adiabatic condition, it is interesting to note that, although $\langle Q\rangle > 0$, the distribution of the generated heat presents negative tails (inset  \autoref{fig:dissipation_soft}).

\section{Conclusions}
In conclusion, in this article we have studied the minimum energy required by reset and switch protocols for a bit encoded by a compressed clamped-clamped graphene ribbon. We presented two protocols that avoid the costly snapping procedure and are specifically designed to reach the minimum heat produced for these logic operations without the need of a fine control of the compression parameter. We stress that it is usually difficult to achieve a precise control of the compression parameter and thus protocols for reset/switch that are robust against variations of such parameter are of great potential interest for practical applications.

\acknowledgments
This work was supported by the European Commission under Grant Agreement No. 318287, LANDAUER and Grant Agreement No. 611004, ICT-Energy.

\end{document}